\documentclass{article}
\usepackage{spconf,amsmath,graphicx,hyperref}
\usepackage{float}
\usepackage{multirow} 
\usepackage{amssymb}
\usepackage{booktabs,tabularx,makecell}

\title{Brain-Grasp: Graph-based Saliency Priors for Improved fMRI-based Visual Brain Decoding}
\name{Mohammad Moradi, Morteza Moradi, Marco Grassia, Giuseppe Mangioni}
\address{
University of Catania, Catania, Italy\\
mohammad.moradi@phd.unict.it
}

\begin{document}
%
\maketitle
\begin{abstract}
Recent progress in brain-guided image generation has improved the quality of fMRI-based reconstructions; however, fundamental challenges remain in preserving object-level structure and semantic fidelity. Many existing approaches overlook the spatial arrangement of salient objects, leading to conceptually inconsistent outputs. We propose a saliency-driven decoding framework that employs graph-informed saliency priors to translate structural cues from brain signals into spatial masks. These masks, together with semantic information extracted from embeddings, condition a diffusion model to guide image regeneration, helping preserve object conformity while maintaining natural scene composition. In contrast to pipelines that invoke multiple diffusion stages, our approach relies on a single frozen model, offering a more lightweight yet effective design. Experiments show that this strategy improves both conceptual alignment and structural similarity to the original stimuli, while also introducing a new direction for efficient, interpretable, and structurally grounded brain decoding.
\end{abstract}
\section{Introduction}
\label{sec:intro}

Visual brain decoding based on fMRI \cite{huang2021fmri} has advanced rapidly with the advent of diffusion-based generative models \cite{cao2024survey} and large vision–language models \cite{zhang2024vision}. Recent methods \cite{scotti2023reconstructing,wang2024mindbridge,ma2025brainclip} achieve substantially higher fidelity and quality in reconstructions, progress enabled by the latest generation of generative techniques. These advances strengthen pipelines in two ways: (i) powerful representations from models such as CLIP \cite{radford2021learning} improve alignment of regions of interest with visual features, yielding richer and more accurate conceptualizations of brain signals, and (ii) diffusion-based generators translate latent representations into outputs that are both meaningful and recognizable, pushing visual brain decoding to levels unattainable only a few years ago.

Despite recent progress, fMRI-based visual brain decoding remains in its early stages and faces critical challenges, particularly in preserving object-level attributes such as saliency, count, and spatial arrangement in reconstructed scenes \cite{moradi2025wrong}. These limitations go beyond pixel fidelity and can negatively impact decision-making, task performance, and perceived realism in downstream applications. Moreover, the decoding pipeline is computationally intensive, motivating the reuse of established components—such as intermediate embeddings from prior work—to reduce cost, support sustainable “green AI,” and enable modular pipelines that allow researchers to focus on addressing these unresolved challenges.

For example, Brain-Optimized Inference (BOI) \cite{kneeland2023brain} refines MindEye \cite{scotti2023reconstructing} reconstructions by iteratively sampling candidate images from a diffusion model and selecting those most consistent with measured fMRI activity, thereby improving perceptual quality and brain–image alignment. Similarly, TROI \cite{wang2025troi} builds on MindEye2 \cite{scotti2024mindeye2} but replaces hand-crafted ROIs with a trainable voxel selection module. Using sparse mask training and learning-rate rewinding, it learns compact, subject-specific voxel masks, enhancing cross-subject decoding efficiency and reconstruction quality with fewer samples.

Following this trend, we introduce Brain-GraSP, a subject-specific fMRI-based visual brain decoding framework that leverages Graph-based Saliency Priors to enhance object conformity and reconstruction accuracy. Brain-GraSP uses the MindEye's precomputed fMRI–CLIP embeddings, from which textual cues are extracted and a graph neural network (GNN) is trained to generate saliency maps directly. These semantic and spatial priors, combined with the embeddings, condition a frozen Stable Diffusion model to produce reconstructions. Our main contributions are: (1) introducing a GNN-based method for deriving saliency maps directly from fMRI–CLIP embeddings, (2) incorporating saliency cues to improve structural fidelity, and (3) generating reconstructions that preserve object-level conformity between stimuli and outputs. Experiments demonstrate competitive reconstruction quality while highlighting saliency as a promising prior for future VBD models.

\begin{figure}[t] 
    \centering
    \includegraphics[width=0.9\columnwidth]{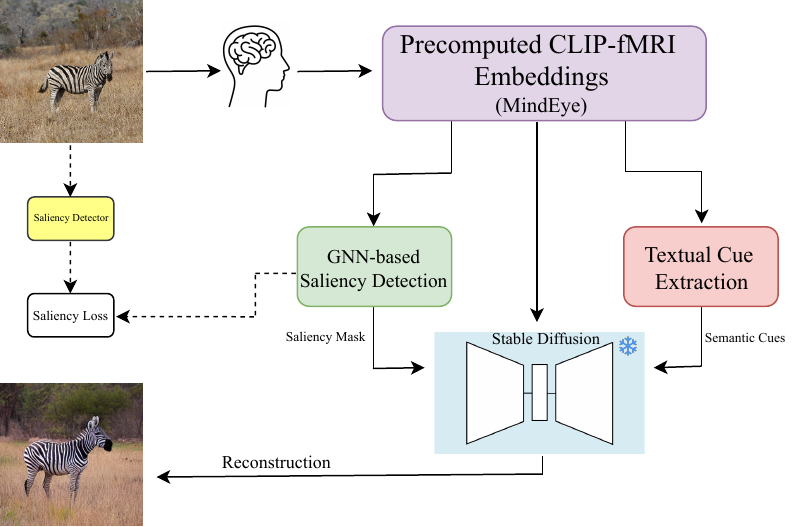}
    \caption{The architecture of Brain-GraSP for fMRI-based visual reconstruction with saliency priors and textual cues. Dashed lines indicate steps used only in saliency detector training, not in the reconstruction pipeline.}
    \label{fig:architecture}
\end{figure}

\section{Methodology}
\subsection{Overview}
We present Brain-GraSP, an fMRI-based decoding framework that integrates saliency and semantic priors into image reconstruction (Figure~\ref{fig:architecture}). Leveraging precomputed CLIP–fMRI embeddings from MindEye, it reduces computational cost while providing rich representations. A GNN predicts saliency from these embeddings, which, together with extracted textual cues, condition a frozen diffusion model to generate reconstructions that preserve object fidelity, mitigate structural inconsistencies, and maintain modularity. As the saliency detector and mapper are optimized per subject, Brain-GraSP is a subject-specific VBD model that enables fine-grained characterization of individual neural representations.

\subsection{fMRI–CLIP Embedding Integration}
We transform fMRI data into multimodal feature representations using the fMRI-CLIP embeddings introduced by \textit{MindEye}, a seminal brain-vision alignment framework. These embeddings capture fine-grained correspondences between neural activity and visual-semantic features, providing a strong initialization for downstream decoding while avoiding the cost of training new encoders. This choice not only reduces computational and energy demands in line with ``green AI'' practices, but also makes the pipeline modular: improvements in upstream embedding models can directly propagate into our framework, which can in turn serve as a test-bed to evaluate the impact of alternative fMRI--CLIP embeddings in the same fMRI-to-image pipeline. Formally, given fMRI measurements $x \in \mathbb{R}^d$, the MindEye encoder outputs CLIP-aligned embeddings $E \in \mathbb{R}^{257 \times 768}$, where $257$ denotes the number of tokens and $768$ is the channel dimension and each token encodes localized visual-semantic information.

\subsection{GNN-based Saliency Modeling}

A central contribution of our framework is the direct estimation of saliency from fMRI--CLIP embeddings using GNNs. Unlike previous approaches that rely on post-hoc analysis of generated images, we target embeddings as the source for saliency because they precede the stochastic and artifact-prone processes of image synthesis. Although embeddings may contain some noise due to imperfect mapping between fMRI signals and visual features, they remain comparatively cleaner and more faithful to the original brain activity. Detecting saliency at this stage ensures that object-level structural information is extracted before additional irregularities from generative models accumulate, thereby improving reliability.
To operationalize this, the embedding tensor 
\(
E \in \mathbb{R}^{257 \times 768}
\) 
is modeled as a graph 
\(
G = (V, \mathcal{E})
\),
where nodes 
\(
v_i \in V
\) 
represent embedding tokens and edges 
\(
(v_i,v_j) \in \mathcal{E}
\) 
encode spatial or semantic proximity. A GNN assigns node-level weights 
\(
\alpha_i
\) 
reflecting importance, which are aggregated to construct a saliency map:
\begin{equation}
S(x,y) = \sigma\!\left(\sum_{i} \alpha_i \, \phi_i(x,y)\right), 
\quad S \in [0,1]^{H \times W},
\end{equation}
where 
\(\phi_i\) 
maps node activations into image space and 
\(\sigma\) 
denotes a logistic normalization. This formulation naturally captures the relational and structural dependencies inherent in visual representations, making GNNs well-suited for saliency inference.

Since this is a novel direction in brain decoding, we designed and implemented three GNN-based saliency detectors: Graph Convolutional Network (GCN) \cite{kipf2016semi}, Graph Attention Network (GAT) \cite{velivckovic2017graph}, and GraphSAGE \cite{hamilton2017inductive}. 
Further details on the training of the saliency detectors are presented in Section~\ref{sec:experiments}. Among the models, GraphSAGE demonstrated the most consistent performance and was chosen as the backbone, with quantitative results reported in the Experiments section.

\subsection{Semantic Cue Extraction}

To complement spatial saliency priors, we extract \textit{semantic textual cues} directly from fMRI-CLIP embeddings. Given an embedding tensor $E \in \mathbb{R}^{257 \times 768}$, we project it into CLIP’s text-aligned space via a mapping function: 
\[
f: \mathbb{R}^{257 \times 768} \rightarrow \mathbb{R}^{d},
\] 
yielding a textual embedding $T = f(E)$. Candidate textual descriptors are then obtained by measuring similarity between $T$ and a vocabulary of CLIP text tokens. This process essentially converts brain-derived embeddings into text-like descriptors that can guide image generation.

Extracting semantics at the embedding level has two advantages. First, embeddings remain closer to the neural source, avoiding distortions introduced during generative sampling. Second, since $E$ is already aligned with CLIP’s multimodal space, $T$ provides language-grounded descriptors directly usable for conditioning. Crucially, deriving both saliency priors and semantic cues from the same embedding ensures compatibility between structural and conceptual priors, reducing conflicts between ``what'' and ``where'' and generating reconstructions that are more coherent in spatial arrangement and semantic fidelity.

\subsection{Saliency-Guided Image Generation}

Image synthesis is performed using a pretrained Stable Diffusion pipeline 
$\mathcal{D}_{\theta}$ integrated with an IP-Adapter module. The IP-Adapter receives fMRI-derived embeddings $E \in \mathbb{R}^{257 \times 768}$ and maps them via a lightweight network $f_{\phi}$ into a conditioning vector $z \in \mathbb{R}^{1024}$, which is L2-normalized and injected into the UNet backbone of $\mathcal{D}_{\theta}$. This mechanism enables conditioning without retraining the diffusion model, ensuring modularity and efficiency.  

Two complementary priors guide the generation: (i) \textbf{saliency masks} 
$S \in [0,1]^{H \times W}$, where $H$ and $W$ denote the height and width, obtained from GNN outputs, which encode spatial structure, and (ii) \textbf{textual cues} $T$, extracted from embeddings, which provide high-level semantic descriptions of the visual scene. Both $S$ and $T$ are incorporated through the IP-Adapter \cite{ye2023ip} and prompt conditioning.  Saliency is applied in three modes: (1) \emph{inpainting}, where $S$ defines regions for localized refinement; (2) \emph{mask-blend}, where foreground $I_{\text{fg}}$ and background $I_{\text{bg}}$ are generated separately and fused as 
$I = S \odot I_{\text{fg}} + (1-S) \odot I_{\text{bg}}$; and 
(3) \emph{ranked generation}, where candidates $\{I_i\}$ are scored by:  
\begin{equation}
\text{score}(I_i) = \lambda_{\text{CLIP}} \, s_{\text{CLIP}}(I_i,T) 
+ \lambda_{\text{mask}} \, s_{\text{mask}}(I_i,S),
\end{equation}
where $s_{\text{CLIP}}$ and $s_{\text{mask}}$ measure semantic alignment (cosine similarity) and structural consistency (IoU/Dice), respectively, and $\lambda_{\text{CLIP}}, \lambda_{\text{mask}}$ are tuned hyperparameters. The highest-scoring candidate is then selected as the final reconstruction.

\section{EXPERIMENTS}
\label{sec:experiments}
To evaluate the performance of \textit{Brain-GraSP}, we followed common best practices in the field as introduced in \cite{scotti2023reconstructing}. Reconstructions were generated from fMRI recordings provided by the benchmark NSD dataset for subjects 1, 2, 5, and 7. 
For a fair comparison, as our GNN-based saliency detector was trained on the last 301 (of 982) images from the NSD test set, we conducted evaluation on the first 681 images, comparing our method against state-of-the-art models (MindEye \cite{scotti2023reconstructing}, MindBridge \cite{wang2024mindbridge} and BOI \cite{kneeland2023brain}) for which reconstructions were either publicly available, provided upon request, or reproduced by us.

We re-evaluated all models on this shared subset using widely adopted metrics, including pixel correlation (PixCorr) for low-level intensity alignment, SSIM for perceptual structural similarity, AlexNet-2/5 \cite{krizhevsky2012imagenet} features for low- and high-level visual representations, CLIP similarity for semantic alignment, Inception Score \cite{szegedy2016rethinking} for image quality and diversity, EfficientNet (EffNet-B) \cite{tan2019efficientnet} features for multi-scale content representation, and SwAV similarity for instance-level consistency in a self-supervised space.

As previously noted, to identify the most effective GNN architecture for detecting saliency directly from embeddings, we evaluated three implementations: GCN, GAT, and GraphSAGE. To train the saliency detectors on the last 301 images of the NSD test set, we employed the EDN model \cite{wu2022edn} to generate saliency maps, which served as ground truth since the NSD dataset does not provide salient object masks. It is worth mentioning that using EDN-generated saliency maps as pseudo–ground truth may introduce bias toward that model’s inductive priors and does not reflect absolute human attention. Nevertheless, due to the absence of human-annotated saliency masks in NSD, EDN is used solely as a proxy reference for relative, cross-method comparison, with the analysis focusing on structural consistency rather than agreement with a single saliency model.

Training was performed using the Adam optimizer with a learning rate of $1 \times 10^{-3}$. The objective function was defined as  
$\mathcal{L} = \mathcal{L}_{\text{wIoU}}(Sal, GT) + \mathcal{L}_{\text{wBCE}}(Sal, GT)$
where $Sal$ is the predicted saliency mask and $GT$ is the ground truth. Here, $\mathcal{L}_{\text{wIoU}}$ denotes the weighted Intersection-over-Union (IoU) loss, and $\mathcal{L}_{\text{wBCE}}$ represents the weighted binary cross-entropy (BCE) loss. As shown in Table~\ref{tab:GNN}, the GraphSAGE implementation achieved the highest average performance across the four subjects. Consequently, the saliency generated by this model was selected as the spatial prior for conditioning Stable Diffusion.

\begin{figure}[t] 
  \centering
  \includegraphics[width=0.55\linewidth]{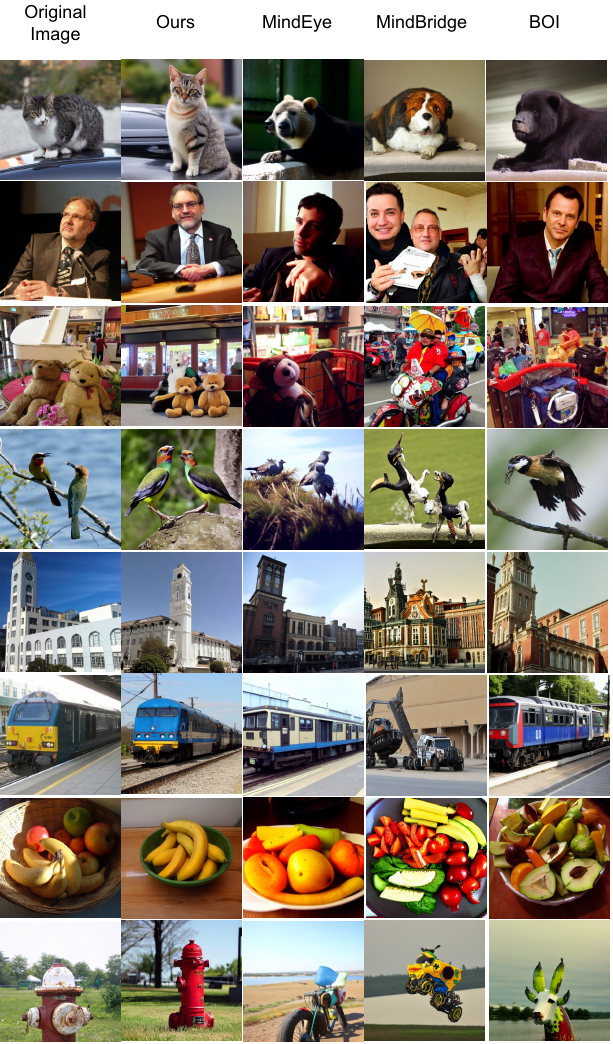} 
  \caption{Qualitative examples of reconstructions by our model, MindEye, MindBridge and BOI.}
  \label{fig:qmap}
\end{figure}

The qualitative comparison of reconstructions from each model is presented in Figure~\ref{fig:qmap}.

\begin{table*}
\centering
\caption{Quantitative evaluation of Brain-GraSP against three state-of-the-art models (top), along with a subject-wise performance breakdown. Highlighted scores indicate better performance.}
\label{tab:main}
\small 
\begin{tabular}{|c|c|c|c|c|c|c|c|c|}
\hline
\textbf{Models} & \textbf{PixCorr} $\uparrow$ & \textbf{SSIM} $\uparrow$ & \textbf{Alex(2)} $\uparrow$ & \textbf{Alex(5)} $\uparrow$ & \textbf{Incep} $\uparrow$ & \textbf{CLIP} $\uparrow$ & \textbf{EffNet-B} $\downarrow$ & \textbf{SwAV} $\downarrow$ \\
\hline
MindBridge & 0.120 & 0.266 & 85.31\% & 93.86\% & 91.94\% & 93.15\% & 0.725 & 0.421  \\ 
\hline
BOI        & 0.260 & 0.268 & 94.38\% & 97.78\% & 93.72\%  & 93.97\% & 0.649 & 0.364 \\ 
\hline
MindEye    & 0.305 & 0.292 & \textbf{95.83\% }& \textbf{97.92\%} & 94.22\%  & 93.45\% & \textbf{0.370} & 0.655 \\ 
\hline
\textbf{Brain-GraSP} & \textbf{0.326} & \textbf{0.299} & 92.79\% & 96.68\% & \textbf{97.85\%} & \textbf{98.80\%} & 0.562 & \textbf{0.315} \\ 
\hline\hline
Subj01 & 0.327 & 0.296 & 93.41\% & 97.46\% & 98.97\% & 99.23\% & 0.535 & 0.306 \\ 
\hline
Subj02 & 0.330 & 0.304 & 93.26\% & 97.22\% & 98.84\% & 99.30\% & 0.533 & 0.306 \\ 
\hline
Subj05 & 0.315 & 0.295 & 90.69\% & 94.82\% & 94.59\% & 97.15\% & 0.645 & 0.346 \\ 
\hline
Subj07 & 0.330 & 0.301 & 93.80\% & 97.22\% & 99.01\% & 99.53\% & 0.533 & 0.303 \\ 
\hline
\end{tabular}
\end{table*}

\section{Result Analysis and Ablation Studies}

According to the performance analysis in Table ~\ref{tab:main} (both in comparison with other models and on a subject-wise basis for our model), the proposed Brain-GraSP demonstrates superior results on most metrics compared to state-of-the-art baselines. The gains are particularly evident in PixCorr, SSIM, Inception, CLIP, and SwAV, which together assess structural fidelity, semantic alignment, and representation-level similarity. These improvements highlight the effectiveness of our strategy that leverages saliency priors and textual cues to enhance object conformity and semantic understanding in the reconstructions. While Brain-GraSP does not achieve the top scores on AlexNet-based features, it consistently outperforms on higher-level metrics, which are more indicative of naturalistic image quality and human-aligned perception.

To ensure a fair comparison, we re-evaluated MindEye, BOI, and MindBridge on a reduced test set of 681 images from NSD and compared their results with the originally reported scores on the full set of 982 images. For most metrics, the re-evaluated scores differed by less than 4\% from the full-set results, confirming that the reduced set is highly representative. This indicates that Brain-GraSP’s performance is not only competitive on the reduced set but can also be strongly extended to the full test set, further underscoring its robustness and promise.

To assess the role of saliency and semantic cues in Brain-GraSP, we conducted ablation studies on Subj01 (Table ~\ref{tab:ablation}). Saliency priors enhance pixel-level fidelity and semantic alignment, while textual cues improve perceptual and representation-based similarity. Using only embeddings leads to a sharp performance drop, highlighting that gains arise not just from embeddings but from how the architecture leverages them. Notably, Brain-GraSP harnesses MindEye’s precomputed fMRI–CLIP embeddings more effectively than the original MindEye pipeline.

Moreover, in selecting an appropriate GNN-based model for detecting saliency from embeddings—an essential component of our pipeline—Table ~\ref{tab:graph} shows that GraphSAGE achieves the best overall performance. Its higher scores across key metrics indicate more accurate saliency localization, stronger structural preservation, and better alignment with salient regions. This advantage comes from its inductive neighborhood aggregation \cite{hamilton2017inductive}, which captures local structure and feature diversity more effectively than GCN or GAT, making it well-suited for modeling the distributed patterns in fMRI–CLIP embeddings.

\begin{table}[t]
\centering
\caption{Performance of GNN-based saliency detection methods across four subjects. (Higher is better for all metrics except MAE)}
\label{tab:ablation}
\scalebox{0.9}{
\begin{tabular}{|l|c|c|c|c|}
\hline
\textbf{Methods}   & \textbf{MAE $\downarrow$} & \textbf{F-max $\uparrow$} & \textbf{E-max $\uparrow$} & \textbf{Fbw $\uparrow$} \\ \hline
GAT       & \textbf{0.1435} & 0.5627 & 0.7611 & 0.4722 \\ \hline
GraphSAGE & 0.1480 & \textbf{0.5813} & \textbf{0.7719} & \textbf{0.4800} \\ \hline
GCN       & 0.1643 & 0.5072 & 0.7206 & 0.4187 \\ \hline
\end{tabular}}
\label{tab:GNN}
\end{table}

\begin{table}[t]
\centering
\caption{Performance comparison of three different GNN-based saliency detection methods.}
\label{tab:graph}
\scalebox{0.82}{
\begin{tabular}{|l|c|c|c|c|}
\hline
\textbf{Methods}   & \textbf{PixCorr $\uparrow$} & \textbf{SSIM $\uparrow$} & \textbf{Alex(2) $\uparrow$} & \textbf{SwAV $\downarrow$} \\ \hline
w/o Sal. Prior       & 0.245 & 0.280 & 89.36\% & 0.318 \\ \hline
w/o Text Cues & 0.314 & 0.248 & 86.05\% & 0.403 \\ \hline
Only Embeds      & 0.110 & 0.206 & 73.32\% & 0.440 \\ \hline
\end{tabular}}
\label{tab:avg_results}
\end{table}

\section{Conclusion}

In this work, we propose Brain-GraSP, an fMRI-based VBD model that incorporates saliency masks and textual cues into Stable Diffusion, achieving superior performance over state-of-the-art baselines. While we follow best practices by reusing precomputed CLIP–fMRI embeddings from a seminal work, the tailored design of our pipeline enables Brain-GraSP to outperform that base model.This highlights opportunities for designing modular pipelines that reuse components to reduce cost and enhance VBD with semantic–spatial cues.

\paragraph*{ACKNOWLEDGMENT:}
The authors acknowledge financial support from PNRR
MUR project PE0000013-FAIR.

\bibliographystyle{IEEEbib}
\bibliography{Main,refs}

\end{document}